# A Distributed k-Secure Sum Protocol for Secure Multi-Party Computations

Rashid Sheikh, Beerendra Kumar, Durgesh Kumar Mishra

**Abstract**—Secure sum computation of private data inputs is an interesting example of Secure Multiparty Computation (SMC) which has attracted many researchers to devise secure protocols with lower probability of data leakage. In this paper, we provide a novel protocol to compute the sum of individual data inputs with zero probability of data leakage when two neighbor parties collude to know the data of a middle party. We break the data block of each party into number of segments and redistribute the segments among parties before the computation. These entire steps create a scenario in which it becomes impossible for semi honest parties to know the private data of some other party.

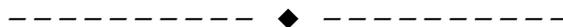

## 1 INTRODUCTION

Due to the tremendous growth of the Internet and huge amount of online transactions taking place over it, opportunities exist for joint computations requiring privacy of the inputs. These computations occur between parties which may or may not have trust in one another. In literature, this subject of information security is called Secure Multiparty Computation (SMC). This type of computation is aimed at privacy of individual inputs and the correctness of the result. Formally, in SMC the parties $P_1, P_2, ..., P_k$ want to compute some common function $f(x_1, x_2, ..., x_k)$ of inputs $x_1, x_2, ..., x_k$ such that a party $P_i$ can know only its own input $x_i$ and the value of the function $f$. The SMC problems use two computation paradigms; ideal model and real model paradigm. In ideal model there exists a Trusted Third Party (TTP) which accepts inputs from all the parties, evaluates the common function and sends result of the computation to the parties. If the TTP is honest, then the parties can know the result only. In real model, there is no third party, instead all the parties agree on some protocol which allows them to evaluate the function while preserving privacy of individual inputs. Consider an example of SMC where two banks cooperatively want to know details about some suspicious customer but no bank is willing to disclose the details of the customer to other bank or even the result of the query. In such situations the SMC solutions are of great significance. A simple and easily understood example of SMC is the secure sum computation where all the parties want to know the sum of their individual data inputs preserving the privacy of inputs [10]. The secure sum protocol proposed by Clifton et al. [10] used randomization method for computing the sum. Parties were arranged in a unidirectional ring. One of the parties is selected as a protocol initiator party which starts the computation by selecting a random number and adds its own data. The sum is then transmitted to the next party. The next party simply adds the data to the received sum and then sends this new sum to the next party. This procedure is repeated until the protocol initiator receives the sum of all the data and the random number. Since the random number is known only to the protocol initiator party, it subtracts the random number from the sum and announces the result to all the parties. The architecture of secure sum protocol is depicted in fig. 1 where parties $P_0, P_1, P_2$ and $P_3$ perform secure sum computation over their data 10, 8, 7 and 15. The protocol initiator party $P_0$ selects a random number $R = 5$ and sends the sum of this random number and its private data to $P_1$. This sum (15) is added with the data of $P_1$ to get the new sum 23 which is sent to $P_2$. This process is repeated until the sum 45 is received by $P_0$. Now, since the random number is only known to $P_0$ it simply subtracts this random number from 45 and broadcasts the final sum as 40 to all the parties. This protocol works well when all the parties are honest in the sense they follow each step agreed in the protocol and never try to know other information except the result of the computation. In this protocol, two adjacent parties to a middle party can cooperate maliciously to know the data of a middle party. We proposed novel protocols in [11, 15, 16] where the probability of data leakage has been reduced by breaking the data block into a fixed number of segments.

In this paper, we propose a secure sum computation protocol with zero probability of data leakage when two adjacent parties to a middle party collude to hack the data of a middle party. Each data block is broken into $k$ segments where $k$ is equal to the number of parties. Then the segments are distributed to other parties before computation. This protocol we call as *dk-Secure Sum* Protocol.

- *Rashid Sheikh and Beerendra Kumar are with the Sri Satya Sai Institute of Science and Technology, Sehore, India,*
- *Durgesh Kumar Mishra is with the Acropolis Institute of Technology and Research, Indore, India.*

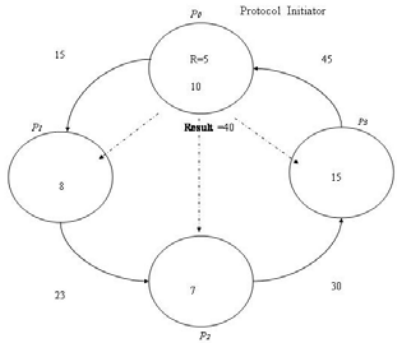

Fig. 1: Secure Sum Protocol [10].

## 2 BACKGROUND

The concept of SMC began in 1982 when Yao proposed and gave solution to his millionaire's problem in which two millionaires wanted to know who was richer without disclosing individual wealth to each other [1]. It was a two-party computation protocol for semi honest parties who follows the protocol but also tries to know something other than the result. The idea was extended to multiparty computation by many researchers [2]. They used circuit evaluation protocols for secure computation. Earlier research focused on theoretical studies. Later, some real life applications emerged like Private Information Retrieval (PIR) [3, 4], Privacy-preserving data mining [5, 6], Privacy-preserving geometric computation [7], Privacy-preserving scientific computation [8], Privacy-preserving statistical analysis [9] etc. A detailed review of SMC research is provided by Du et al. in [12] where they developed a framework for problem discovery and converting normal problem to SMC problem. A review of SMC with special focus on telecommunication systems is given by Oleshchuk *et al.* in [13]. Anonymity enabled SMC was proposed by Mishra et al. in [14] where the identities of the parties are made ambiguous for achieving privacy.

Our proposed protocol is motivated by the work of Clifton *et al.* [10] where they proposed a toolkit of components for solution to SMC problems. One of the components of this toolkit is the secure sum computation. This component is used in many distributed data mining applications where many physically distributed sites compute sum of values from individual sites. The secure sum protocol proposed in [10] used random numbers for privacy of individual data inputs. The weakness of this scheme is that any two parties $P_{i-1}$ and $P_{i+1}$ can cooperate to know the secret data of party $P_i$ by performing only one computation. We proposed *k-Secure Sum* Protocol and *Extended k-Secure Sum* Protocol in [11] where we significantly reduced the probability of data leakage by breaking the data block of individual party in number of segments. The probability of data leakage decreases as we increase the number of segments in a data block.

We proposed novel protocols for reducing probability of data leakage to zero by changing the positions of the parties in the ring [15, 16].

## 3 PROPOSED ARCHITECTURE AND PROTOCOL DESCRIPTION

Figure 2 shows the proposed protocol architecture for *dk-Secure Sum* Protocol before redistribution of segments. Cooperating parties $P_1$, $P_2$, …, $P_k$ are arranged in a ring for computation purpose. Each party itself breaks the data block into $k$ segments. There also exist distribution paths for distributing the data segments to other parties before computation. The computation path is shown with a solid line but the distribution path is shown with the dashed line. The communication lines between parties are assumed to be secure.

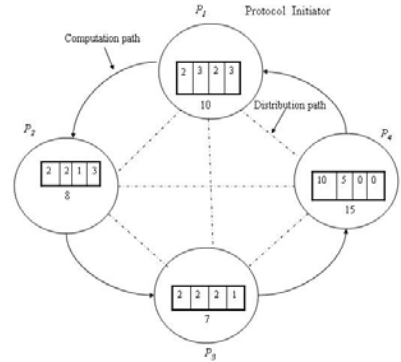

Fig 2: Architecture of *dk-Secure Sum* Protocol before redistribution.

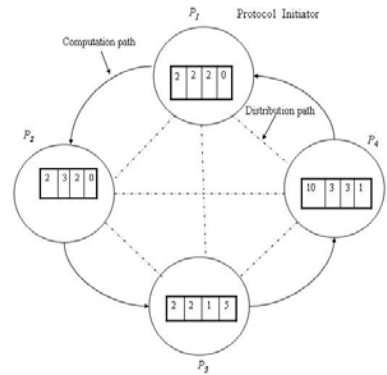

Fig 3: Snapshot of *dk-Secure Sum* Protocol after redistribution.

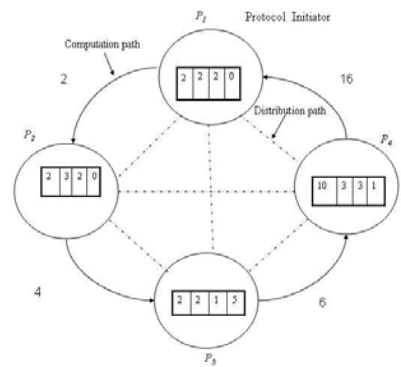

Fig. 4: Snapshot of *dk-Secure Sum* Protocol for first round of segment computation

### 3.1 Informal Description of *dk-Secure Sum* Protocol

Assume, $P_1$, $P_2$, ..., $P_k$ are $k$ parties involved in cooperative secure sum computation where each party is capable of breaking its data block into a fixed number of segments such that the sum of all the segments is equal to the value of the data block of that party. In proposed protocol number of segments in a data block is kept equal to the number of parties. The values of the segments are randomly selected by the party and it a secret of the party. If $k$ be the number of segments (which is same as the number of parties) then in this scheme each party holds any one segment with it and $k$-1 segments are sent to $k$-1 parties, one to each of the parties. Thus at the end of this redistribution each of the parties holds k segments in which only one segment belongs to the party and other segments belong to remaining parties, one from each. A snapshot of four party secure sum computations after distribution of segments is shown in fig 3. Now, *k-Secure Sum* Protocol [11] can be applied to get the sum of all the segments. In this protocol, one of the parties is unanimously selected as the protocol initiator party which starts the computation by sending the data segment to the next party in the ring. The receiving party adds its data segment to the received partial sum and transmits its result to the next party in the ring. This process is repeated until all the segments of all the parties are added and the sum is announced by the protocol initiator party. Snapshot of *dk-Secure Sum* Protocol for first round of segment computation is shown in the fig 4.

Now even if two adjacent parties maliciously cooperate to know the data of a middle party they will be able to know only those $k$ segments of a party which belong to every party. The sum of these segments is a garbage value and thus worthless for the hacker party.

### 3.2 Formal Description of *dk-Secure Sum* Protocol

Our protocol is based on the architecture shown in fig 2 and fig 3. There are $k$ parties $P_1$, $P_2$, ..., $P_k$ involved in cooperative secure sum computation. Each party $P_i$ has its secret input $x_i$ and all the parties are interested to know the sum $\sum x_i$ for $i = 1$ to $k$.

Algorithm: *dk-Secure Sum*
1. Define $P_1$, $P_2$, ..., $P_k$ as $k$ parties.
2. Assume these parties have secret inputs $x_1, x_2, ..., x_k$.
3. Each party $P_i$ breaks its data $x_i$ into $k$ segments $d_{i1}$, $d_{i2}$, ..., $d_{ik}$ such that $\sum d_{ij} = x_i$ for $j = 1$ to $k$.
4. Each party keeps any one segment with it and distributes $k$-1 segments to other parties such that one segment is distributed to one party.
5. Each party reshuffles the received segments randomly.
6. Assume $rc = k$ and $S_{ij} = 0$,
/* $S_{ij}$ is partial sum and rc is round counter */
7. while $rc$!=0
   begin
   for $j = 0$ to $k$-1
   for $i = 0$ to $k$-1
   $P_i$ sends $S_{ij} = d_{ij} + S_{ij}$ to $P_{(i+1) \mod k}$
   $rc = rc - 1$
   end
8. The protocol initiator party broadcasts sum $S_{ij}$ to all the parties.
9. End of algorithm.

### 3.3 Performance Analysis of *dk-Secure Sum* Protocol

The *dk-Secure Sum* Protocol performs satisfactorily due to many reasons. The segmentation of the data block of a party is done by the party in its own way. Secondly, the party randomly selects any one of the segments and keeps with it. Remaining segments are distributed to other parties randomly. At the end of such a distribution each party holds equal number of segments. Each party holds one segment of its own and one segment from each of the parties. Now, if $P_{i+1}$ and $P_{i-1}$ collude to know $x_i$ which is the secret of $P_i$ then they will be able to know only these segments. The sum of these segments is garbage for the colluding parties. The protocol may not work correctly for three parties because if any two parties become corrupt then two segments directly received from the third party can be shared and the third segment can be attacked during computation. But as the number of segments increases beyond three then other segments become ambiguous for the corrupt parties as there is know way to know all the segments of a middle party. Thus the probability of data leakage for four or more parties is zero when majority of parties are honest. This is an appreciable improvement over previous protocols [11] where the probability of data leakage by two colluding parties was $P_{1k} = (2/n\text{-}1)^k$ where $k$ is the number of segments and n is the number of parties.

Now, to calculate the communication complexity of the *dk-Secure* algorithm, we know that each party distributes $k$-1 segments to other parties. Therefore $k$ ($k$-1) communications are done during distribution. During computation each party sends partial sum of the segments to other party in the ring. Thus in each round of computation $k$ communications are done. For $k$ such rounds $k^2$ communications are done. Therefore total number of communications is given in (1).

$C(k) = k(k\text{-}1) + k^2$

$C(k) = 2k^2 - k$ (1)

Now for getting the computation complexity $S(k)$ we see that in each round $k$ computations are done. Thus for $k$ such rounds $k^2$ computations are done which is given in (2).

$S(k) = k^2$ (2)

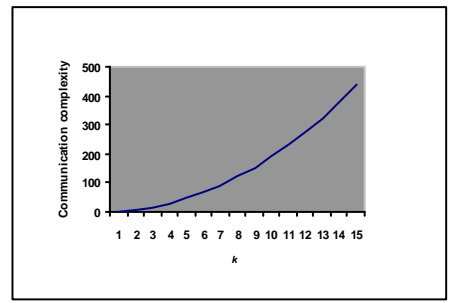

Fig. 5: Communication Complexity as a function of no. of parties in *dk-Secure Sum* protocol

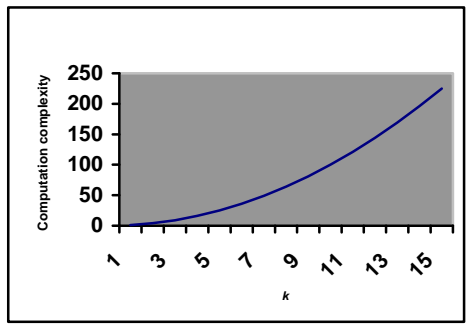

Fig. 6: Computation Complexity as a function of no. of parties in *dk-Secure Sum* protocol.

Figures 5 and 6 depict the communication and the computation complexity with the number of parties respectively. Thus we show that the communication and the computation complexity both are $O(k^2)$.

## 4 CONCLUSION AND FUTURE SCOPE

In this paper, we propose *dk-Secure Sum* Protocol for computation of sum of individual parties preserving privacy of their inputs. The protocol allows parties to break their data inputs into segments and distributing these segments among parties before computation. It provides zero probability for two colluding neighbors when they want to attack data of a middle party. This is an appreciable improvement over previous protocols available in the literature. Further efforts can be made to reduce the computation and the communication complexity preserving the property of zero hacking.

## Authors Profile

### *Rashid Sheikh*

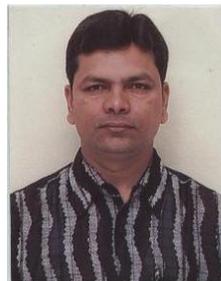

**Biography**: Rashid sheikh has received B.E.(Bachelor of Engineering ) degree in Electronics and Telecommunication Engineering from Shri Govindram Seksaria Institute of Technology and Science, Indore, M.P., India in 1994. He has 15 years of teaching experience. His subjects of interest include Computer Architecture, Computer Networking, Digital Computer Electronics, Operating Systems and Assembly Language Programming. He has published several research papers in National/International Conferences and Journals. His research areas are Secure Multiparty Computation and Mobile Ad hoc Networks. He is the author of ten books on Computer Organization and Architecture.

### *Beerendra Kumar*
Ph. +91 9770435336

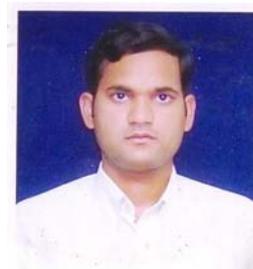

Beerendra Kumar has received B.Tech. (Bachelor of Technology) degree in Computer Science and Information Technology from Institute of Engineering and Technology, Rohilkhand University, Bareilly (U.P), India in 2006. He has completed M.Tech. (Master of Technology) in Computer Science from SCS, Devi Ahilya University, Indore, India in 2008. He has two years of teaching experience. His subjects of interest include Computer Networking,

Theory of Computer Science, Data Mining, Operating Systems and Analysis & Design of Algorithms. He has published three research papers in national conferences and three research papers in international journals. His research areas are Computer Networks, Data Mining, Secure Multiparty Computations and Neural Networks.

### *Dr. Durgesh Kumar Mishra*

Professor (CSE) and Dean (R&D),
Acropolis Institute of Technology and Research, Indore, MP, India,
Ph - +91 9826047547, +91-731-4730038

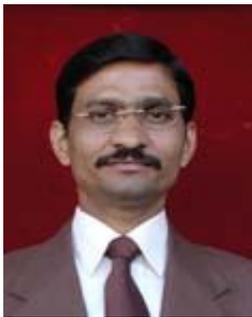

**Biography:** Dr. Durgesh Kumar Mishra has received M.Tech. degree in Computer Science from DAVV, Indore in 1994 and PhD degree in Computer Engineering in 2008. Presently he is working as Professor (CSE) and Dean (R&D) in Acropolis Institute of Technology and Research, Indore, MP, India. He is having around 20 Yrs of teaching experience and more than 5 Yrs of research experience. He has completed his research work with Dr. M. Chandwani, Director, IET-DAVV Indore, MP, India on Secure Multi- Party Computation. He has published more than 65 papers in refereed International/National Journal and Conference including IEEE, ACM. He is a senior member of IEEE, Chairman IEEE Computer Society, Bombay Section and Vice Chairman IEEE MP-Subsection, India. Dr. Mishra has delivered tutorials in IEEE International conferences in India as well as in other countries. He is the programme committee member of several International conferences. He visited and delivered invited talks in Taiwan, Bangladesh, USA, UK, etc. on Secure Multi-Party Computation of Information Security. He is an author of one book. He is reviewer of three International Journals of Information Security. He is a Chief Editor of *Journal of Technology and Engineering Sciences*. He has been a consultant to industries and Government organizations like Sales tax and Labor Department of Government of Madhya Pradesh, India.